 \documentclass[final,5p,times,twocolumn]{elsarticle}

\usepackage{hyperref}

\graphicspath{{images/}}

\journal{Nucl. Instrum. Methods Phys. Res. A}

\begin{document}

\begin{frontmatter}

\title{Minimal material, maximum coverage:\\ Silicon Tracking System for high-occupancy conditions }

\cortext[correspondingauthor]{Corresponding author}
\author[1]{M.~Teklishyn\corref{correspondingauthor}}
\ead{m.teklishyn@gsi.de}
\author[1,2]{L.\,M.~Collazo~S\'anchez}
\author[1]{U.~Frankenfeld}
\author[1]{J.\,M.~Heuser}
\author[3]{O.~Kshyvanskyi}
\author[1]{J.~Lehnert}
\author[1,2]{D.\,A.~Ram\'{\i}rez~Zaldivar}
\author[1,2]{D.~Rodr\'{\i}guez~Garc\'{e}s}
\author[1]{A.~Rodr\'{\i}guez~Rodr\'{\i}guez}
\author[1]{C.\,J.~Schmidt}
\author[1,2,4]{P.~Semeniuk}
\author[1,2]{M.~Shiroya}
\author[1,5]{A.~Sharma}
\author[1,2]{A.~Toia}
\author[1]{O.~Vasylyev}
\author[]{\mbox{for the CBM Collaboration}}

\address[1]{GSI Helmholtzzentrum für Schwerionenforschung GmbH, Planckstraße 1, 64291 Darmstadt, Germany}
\address[2]{Goethe University Frankfurt, Max-von-Laue-Straße 1, 60438 Frankfurt am Main, Germany}
\address[3]{Institute for Nuclear Research, Prospect Nauky 47, 03028 Kyiv, Ukraine}
\address[4]{AGH University of Krakow, Mickiewicza 30, 30-059 Kraków, Poland}
\address[5]{Aligarh Muslim University, Aligarh 202002, India}

\begin{abstract}
Silicon strip sensors have long been a reliable technology for particle detection. Here, we push the limits of silicon tracking detectors by targeting an unprecedentedly low material budget of 2\%–7\%~$X_0$ in an 8-layer 4~m$^2$ detector designed for high-occupancy environments ($\leq$ 10 MHz/cm$^2$).

To achieve this, we employ Double-Sided Double Metal (DSDM) silicon microstrip sensors, coupled with readout electronics capable of precise timing and energy measurements. These 320~$\mu$m thick sensors, featuring $2\times1024$ channels with a 58~$\mu$m pitch, are connected via ultra-lightweight aluminium-polyimide microcables for signal transmission and integrated with a custom SMX readout ASIC, operating in free-streaming mode. This system enables the simultaneous measurement of time ($\Delta t \simeq 5$~ns) and charge deposition (0.1–100~fC), significantly enhancing the detector’s capacity for high-precision track reconstruction in high-occupancy and harsh radiation field environments.

The primary application of this technology is the Silicon Tracking System (STS) for the CBM experiment, with additional potential in projects like the J-PARC E16 experiment and future uses in medical physics, such as advanced imaging telescopes. In this contribution, we present the current status of CBM STS construction, with almost one-third of the modules produced and tested. We also discuss immediate applications and explore promising prospects in both scientific and medical fields.
\end{abstract}

\begin{keyword}
Silicon strip detectors \sep heavy-ion physics \sep high-occupancy detectors \sep free-streaming readout \sep detector integration \sep particle tracking
\end{keyword}

\end{frontmatter}

\section{Silicon Tracking System of the CBM experiment}

The Compressed Baryonic Matter (CBM) experiment at the upcoming FAIR facility in Darmstadt, Germany, is designed to investigate strongly interacting matter under extreme density conditions. Its primary objectives include analysing differential cross-sections and flows of multi-strange particles, studying higher-order fluctuations, and examining dilepton spectra to explore the caloric curve of dense nuclear matter.  These and other observations require high statistics, requiring a beam-target interaction rate of $\leq 10\,\mathrm{MHz}$ with kinetic energies of heavy-ion and proton beams up to $11\,A\mathrm{GeV}$ and $29\,\mathrm{GeV}$, respectively \cite{Agarwal2023, Senger2024}.

The Silicon Tracking System (STS) is a key tracking detector in the CBM experiment, designed to tackle the unique challenges of heavy-ion physics. It operates in an extremely dense environment, handling about thousand charged particle tracks per typical Au-Au collision. The SIS-100 accelerator will deliver continuous high-intensity beams, requiring CBM detectors to function in free-streaming mode. Therefore, the STS will provide timing information to enable on-line event detection, reconstruction and selection \cite{Heuser2024}.

The STS will be housed within a gas-tight thermal enclosure, positioned inside the aperture of the $1\,\mathrm{Tm}$ superconducting dipole magnet, at a distance of $(0.3$–$1.0)\,\mathrm{m}$ downstream of the vacuum chamber, which hosts the target and the Micro-Vertex Detector (MVD) based on Monolithic Active Pixel Sensor (MAPS) technology featuring MIMOSIS chip \cite{Deveaux2025}. These two silicon tracking detectors of CBM share a high degree of mechanical integration, as well as common services and cooling infrastructure \cite{Matejcek2024}. Together with the MVD, the STS plays a crucial role in momentum measurement and vertex reconstruction for complex cascade decays. In particular, CBM performance simulations have demonstrated the ability to reconstruct decays with neutral products, {\it i.\,e.}  $\Sigma^\pm\to n \pi^\pm$ without direct detection of the neutron by identifying the kink between the mother and daughter particle tracks \cite{Senger2024}. Due to the necessity of reconstructing the soft-momentum reaction products, the spatial precision of the STS and MVD is primarily limited by multiple scattering. This constraint significantly influences their mechanical design and integration approaches.

\section{CBM STS integration concept}

\begin{figure}[h]\centering
 \includegraphics[width=.8\linewidth]{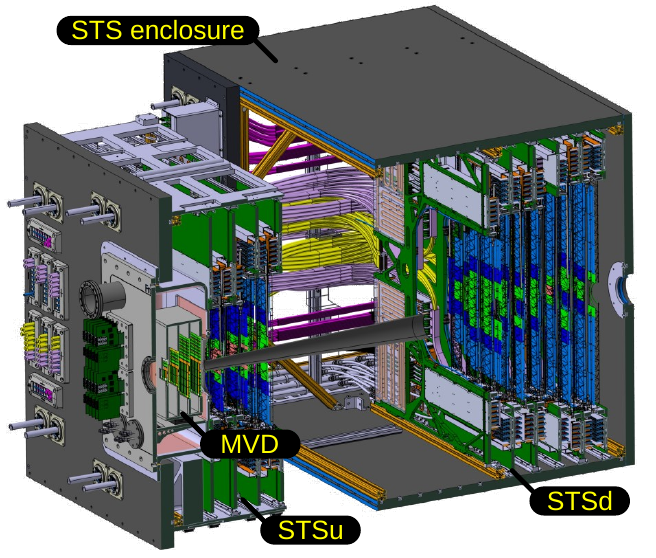}
 \caption{CAD rendering of the STS enclosure cutout, showing the MVD and STSu subdetectors shifted upstream, along with the STSd subdetector.}
 \label{fig:sts-box}
\end{figure}

\begin{figure}[h]\centering
 \includegraphics[width=.49\linewidth]{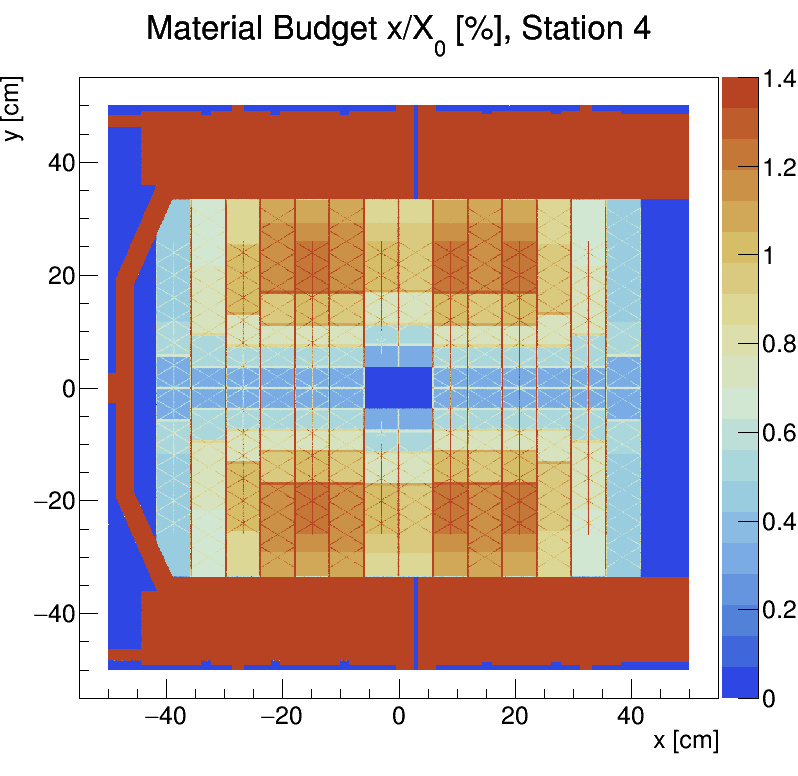}~
 \includegraphics[width=.49\linewidth]{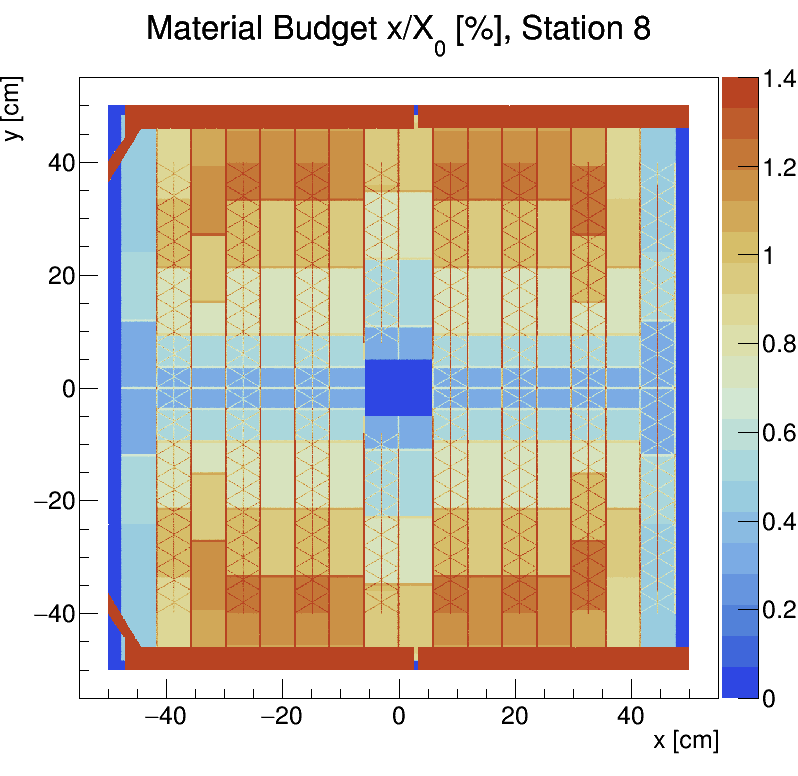}
 \caption{Material budget of the first and the last stations of the CBM STSd.}
 \label{fig:sts-material}
\end{figure}

The STS setup consists of 876 Double-Sided Double-Metal (DSDM) silicon micro-strip modules, arranged in eight tracking stations physically separated in three upstream (STSu) and five downstream (STSd) stations (see Fig.\,\ref{fig:sts-box}). These STS modules incorporate sensors of various sizes, ranging from $22\times62\,\mathrm{mm^2}$ to $124\times62\,\mathrm{mm^2}$, contributing to a total silicon area exceeding $4\,\mathrm{m^2}$ \cite{Teklishyn2024}. Unlike most silicon tracking detectors in particle physics, which position their readout electronics as close as possible to the silicon sensor, the STS employs a challenging technique that utilizes ultra-thin aluminum-polyimide microcables to transmit an unamplified analog signal. This approach allows the readout and powering electronics to be positioned outside the detector aperture, which spans approximately $2.5^\circ - 25.0^\circ$, thereby achieving a material budget of about $0.3\%\,X_0 - 1.4\%\,X_0$ per station as it is shown in Fig.\,\ref{fig:sts-material}.

\begin{figure}[h]\centering
 \includegraphics[width=.40\linewidth]{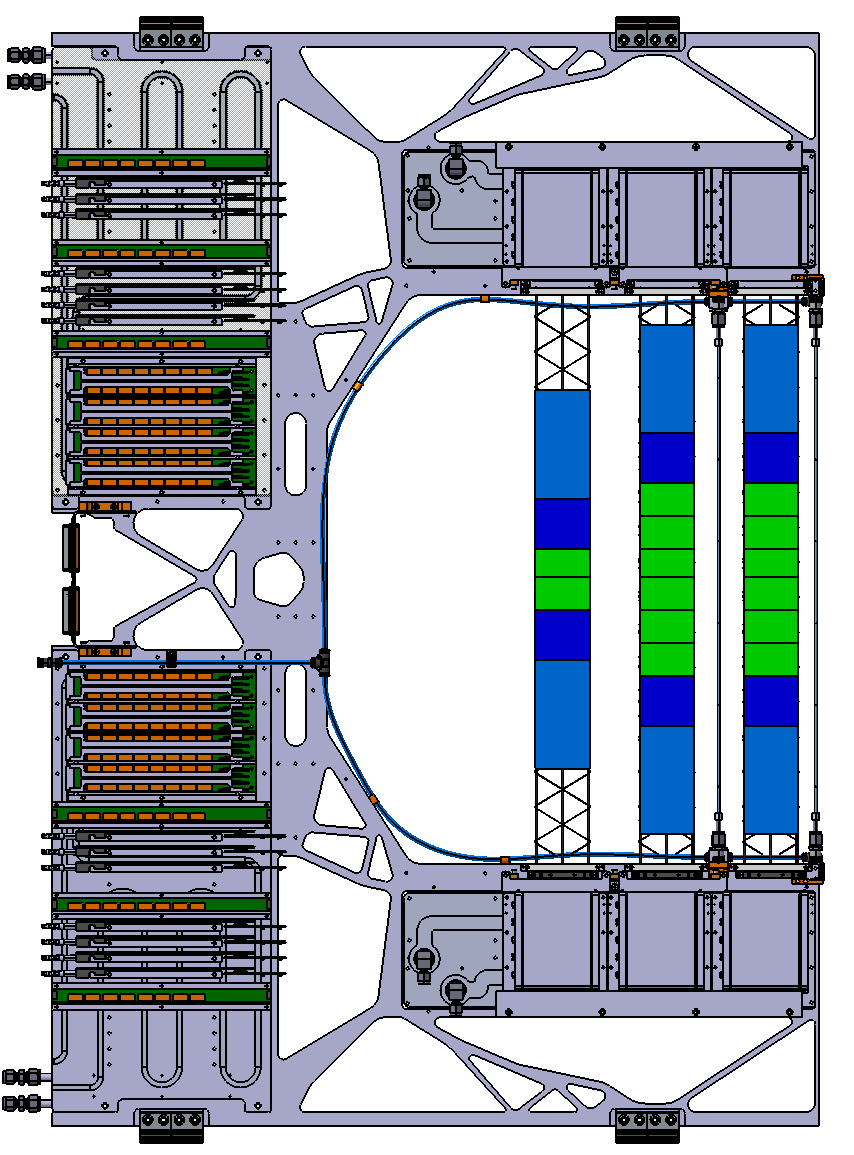}~~~
 \includegraphics[width=.41\linewidth]{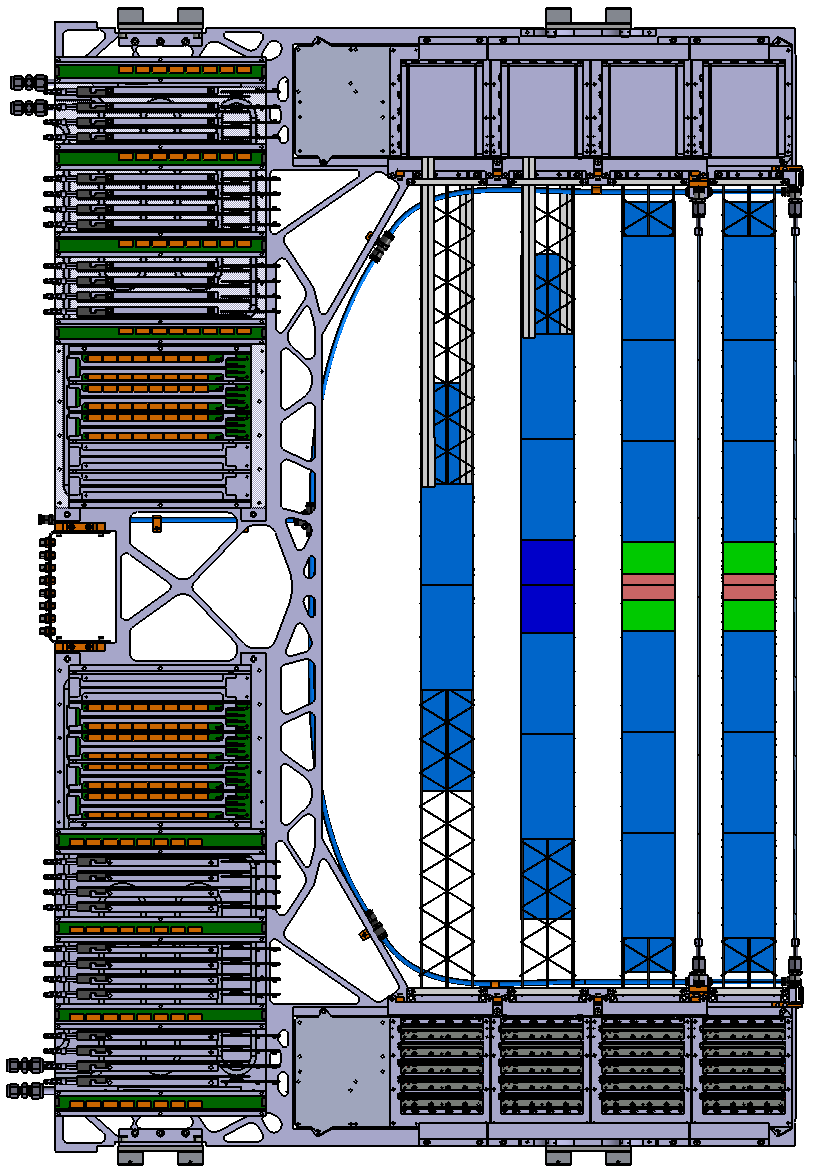}
 \caption{Half-Unit 3 and Half-Unit 7 of the CBM STS.}
 \label{fig:hu}
\end{figure}

The mechanical integration of the STS detector relies on lightweight carbon-fiber support structures (Ladders), each accommodating up to ten modules. Along with the readout and powering electronics, cables, and other services, these ladders are mounted on both sides of aluminum C-shaped support structures, forming Half-Units, as illustrated in Fig.~\ref{fig:hu}. Sensors in adjacent Half-Units create overlapping layers, forming the tracking stations.

Although this design offers significant advantages for detector performance, it also introduces challenges in integration, noise suppression, and ensuring long-term operational stability. In the following sections, we will detail the strategies and methodologies implemented to address these challenges.

\section{Design and features of the silicon micro-strip module}
\paragraph{Module layout}

\begin{figure}[h]\centering
 \includegraphics[width=1\linewidth]{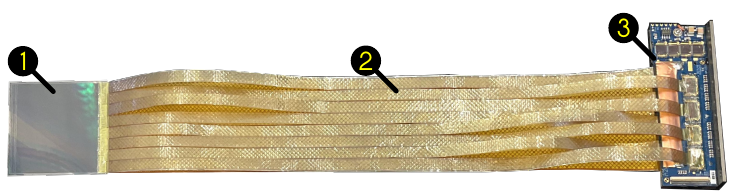}
\caption{STS module featuring (1) a $62\times 62\,\mathrm{mm^2}$ sensor, (2) a stack of 32 microcables, and (3) a two-FEB assembly mounted on an aluminum cooling shelf. The bottom row of four SMX chips is visible on the n-side FEB.}
 \label{fig:module}
\end{figure}

An STS module consists of a DSDM silicon micro-strip sensor with $2\times1024$ strips, all connected via ultra-thin, 64-line aluminum-polyimide microcables to two Front-End Boards (FEBs). Each FEB hosts eight custom-built SMX readout chips. A partially assembled module is shown in Fig.~\ref{fig:module}. Further details on the module layout, prototyping, and performance evaluation can be found in Refs.~\cite{Rodriguez2024, Teklishyn2024}.

\begin{figure}[h]\centering
 \includegraphics[width=.6\linewidth]{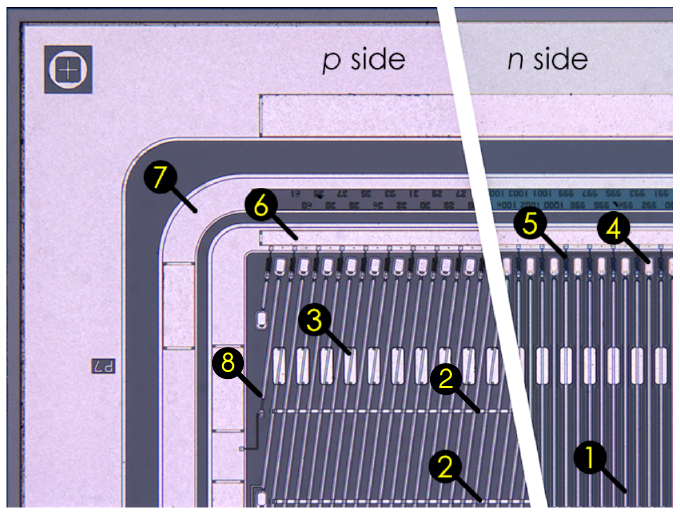}
 \caption{Layouts of the CBM silicon microstrip sensors. Labels indicate:
(1) aluminum readout strips (1st metal layer),
(2) interconnecting lines for Z-strips (2nd metal layer),
(3) AC pads,
(4) DC pads,
(5) bias resistors,
(6) bias ring,
(7) guard ring,
(8) disconnected corner strips
(from Ref.\cite{Panasenko2022}, modified).}
 \label{fig:sensor}
\end{figure}

\paragraph{Silicon micro-strip sensor}
Each STS module is equipped with a $320 \,\mathrm{\mu m} \pm15 \,\mathrm{\mu m}$ thick $p^{++}-n^{+}-n^{++}$ silicon sensor, manufactured by Hamamatsu Photonics (Japan) \cite{hpk}. The strip pitch on both sides is $58\,\mathrm{\mu m}$, with p-side strips tilted at $7.5^\circ$ as it is shown in Fig.\ref{fig:sensor} to enable 2D position reconstruction. The corner strips are interconnected thought the second metalisation layer (Z-strips). The resulting total strip capacitance for n and p-side is  $C_n = 1.02 \pm 0.02 \,\mathrm{pF/cm}$ and $C_p = 1.33 \pm 0.02 \,\mathrm{pF/cm}$ with additional contribution of $15 \,\mathrm{pF}$ for Z-strips \cite{Panasenko2022}.

\begin{figure}[h]\centering
 \includegraphics[width=1\linewidth]{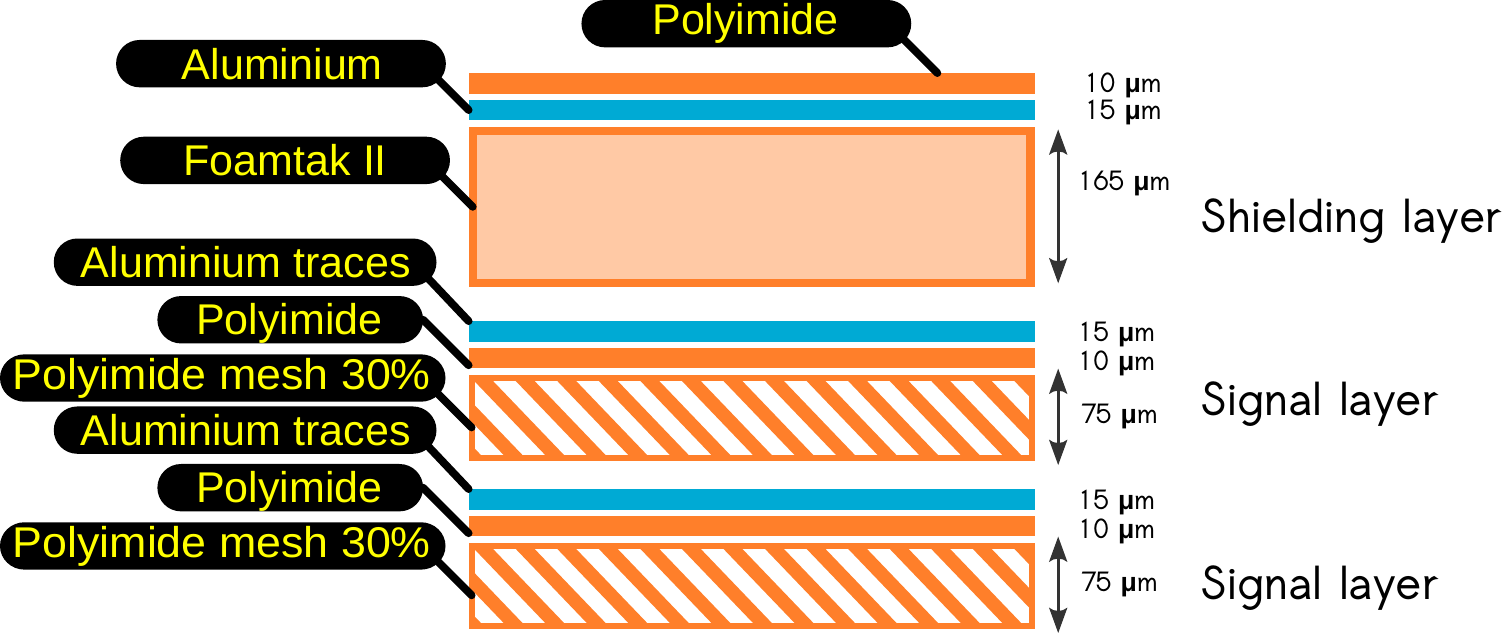}
\caption{Schematics of the set of upper layers of signal micro cables; bottom layers are arranged symmetrically with the shielding underneath.}
 \label{fig:cables}
\end{figure}

\paragraph{Aluminium-polyimide micro cable}
Each SMX chip is connected to the corresponding 128 sensor strips via two microcables, with one handling odd channels and the other even channels (see Fig.\,\ref{fig:cables}). Two common shielding layers cover the entire n-side and p-side of the signal microcable stack, providing electromagnetic protection and noise suppression.

Microcables are delicate and fragile components that require careful handling. It has been observed that the Effective Noise Charge (ENC) of the module is highly sensitive to the precise arrangement of individual microcables within the stack. Wiggling, twisting, or even slight misalignment can introduce excessive noise in the affected channels. The total capacitance of a given trace in the microcable stack was measured to be approximately $C_{{tot}} = 0.38 \pm 0.02\,\mathrm{pF/cm}$. Numerical simulations using the ANSYS package yield closely matching results \cite{Selyuzhenkov:201318}.

\paragraph{SMX read-out chip}

The STS detector utilizes a custom-designed readout chip, STS-MUCH-XYTER (SMX) v2.2, which enables simultaneous free-streaming readout of 128 channels. Each channel features two analog processing paths for the concurrent measurement of amplitude and time. The 5-bit flash analog-to-digital converter (ADC) provides a dynamic range of $<\! 94\,\mathrm{ke}$ ($<\! 625\,\mathrm{ke}$ in low-gain mode), while the 14-bit time-to-digital converter (TDC) operates with a clock period of $3.125\,\mathrm{ns}$ \cite{Kasinski2018}. This capability enables the STS to provide 5D tracking information, simultaneously measuring the interaction position, time, and deposited charge within the silicon sensor \cite{Teklishyn2024}.
The intrinsic equivalent noise charge of an SMX channel has been measured to be approximately $0.35\,\mathrm{ke}$ \cite{Rodriguez2024} while the total noise of the detector depends on the capacitive load.

\paragraph{Layout of FEB8, powering and grounding of the module}

\begin{figure}[h]\centering
 \includegraphics[width=.9\linewidth]{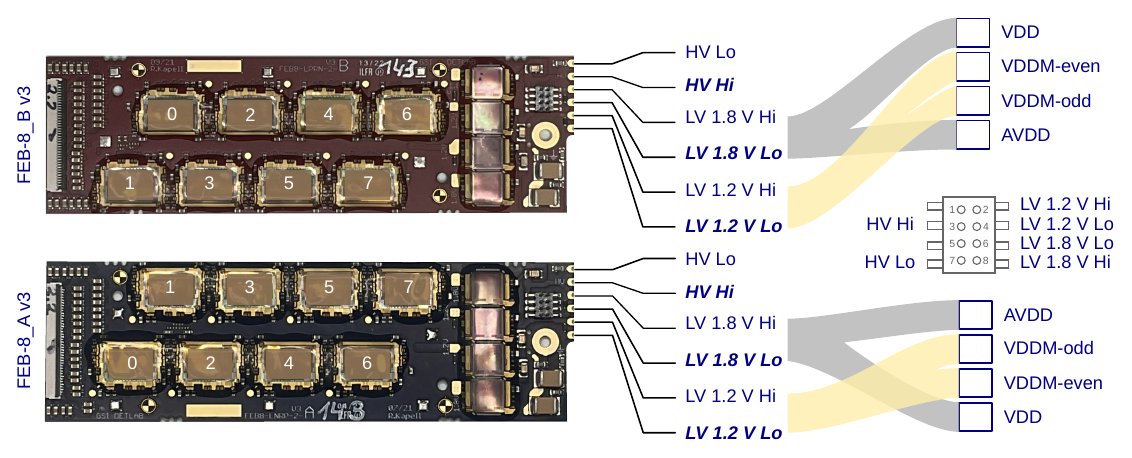}
\caption{Layout and voltages of two flavours of STS FEB8.}
 \label{fig:feb8}
\end{figure}

The schematic of the STS front-end boards, which host eight SMX chips each with two active uplinks, is shown in Fig.~\ref{fig:feb8}. Two powering potentials are provided through four custom-built low-noise, low-dropout (LDO) voltage regulators developed and produced by Semi-Conductor Laboratory (India) \cite{scl2024}, delivering nominal outputs of $1.8\,\mathrm{V}$ and $1.2\,\mathrm{V}$.

\begin{figure}[h]\centering
 \includegraphics[width=.6\linewidth]{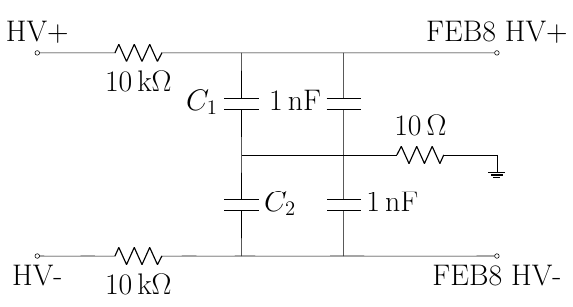}
\caption{Conceptual grounding schematics of the module. In the recent FEB design the $C_{1,\,2}$ are implemented by two sequential  $2\,\mathrm{\mu F}$ SMD capacitors.}
 \label{fig:ground}
\end{figure}

The inverse bias voltage required for the operation of the silicon sensor is supplied through the ground potentials of the corresponding FEB8. It features a symmetric implementation, meaning that the front-end electronics operate under high voltage. The simplified detector grounding schematics, featuring on-board RC filter, is shown in Fig.\ref{fig:ground}.

\begin{figure}[h]\centering
 \includegraphics[width=.9\linewidth]{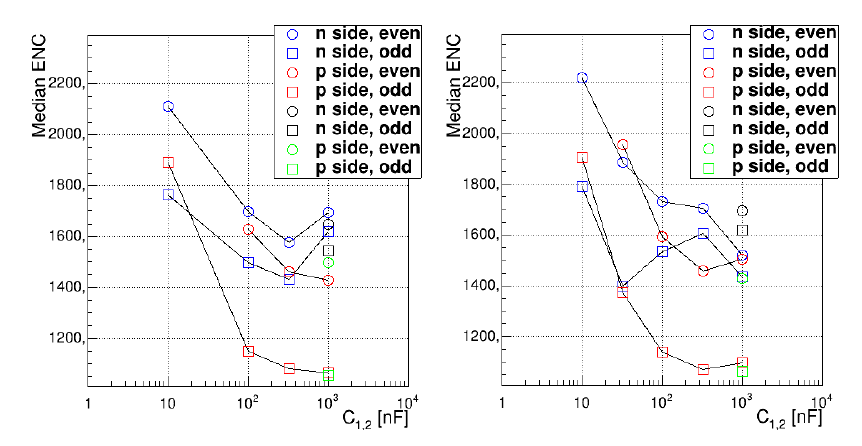}
\caption{The median equivalent noise charge in electrons of the SMX timing channel for odd and even channels of the STS module depending on the stabilising capacitor values \cite{Kshyvanskyi2022}. }
 \label{fig:cscan}
\end{figure}

It was observed that large stabilizing capacitors are crucial for ground stabilization and ensuring low-noise detector operation. Fig.~\ref{fig:cscan} illustrates the noise dependence on the nominal capacitor value.

\section{Performance of the series production modules}
The CBM STS detector is currently in the mass production stage. Over one-third of the detector modules have been assembled and evaluated, providing an initial insight into the anticipated detector performance.

\begin{figure}[h]\centering
  \includegraphics[width=.9\linewidth]{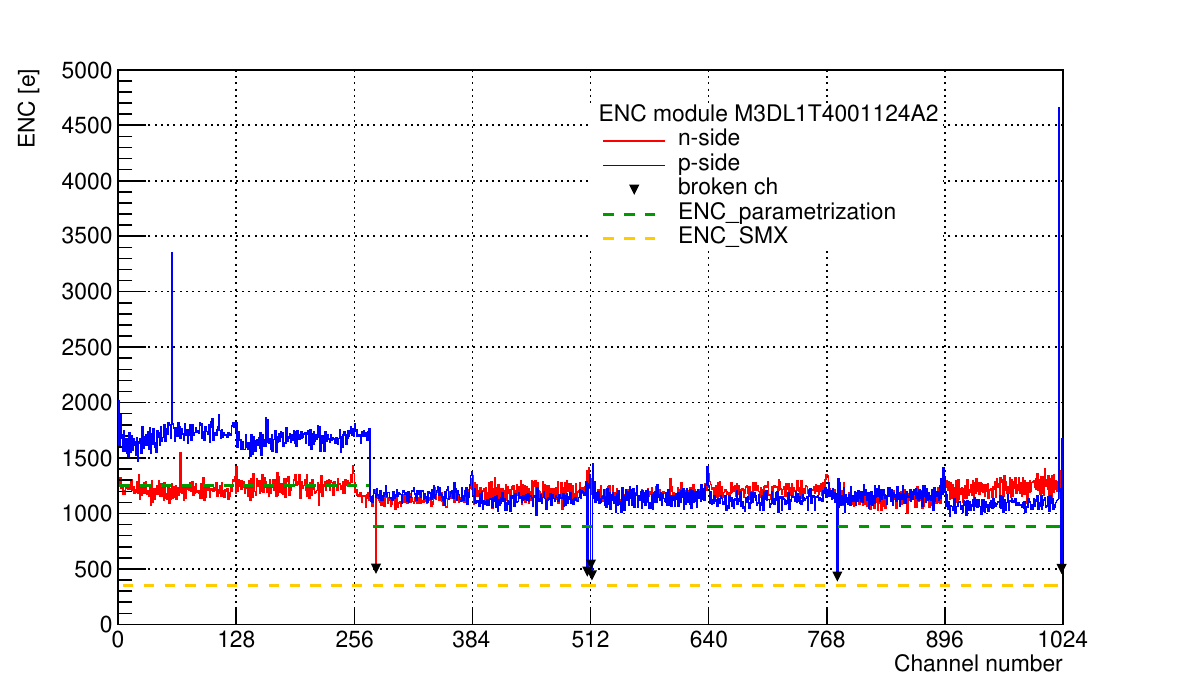}
\caption{Equivalent noise charge for a selected STS module. }
 \label{fig:enc}
\end{figure}

Each module undergoes a calibration procedure for its flash ADC, followed by noise evaluation through pulse-scan analysis. The details of this procedure are provided in Ref.~\cite{Rodriguez2024}. Examples of the resulting plots are shown in Fig.\ref{fig:enc}.

\begin{figure}[h]\centering
 \includegraphics[width=\linewidth]{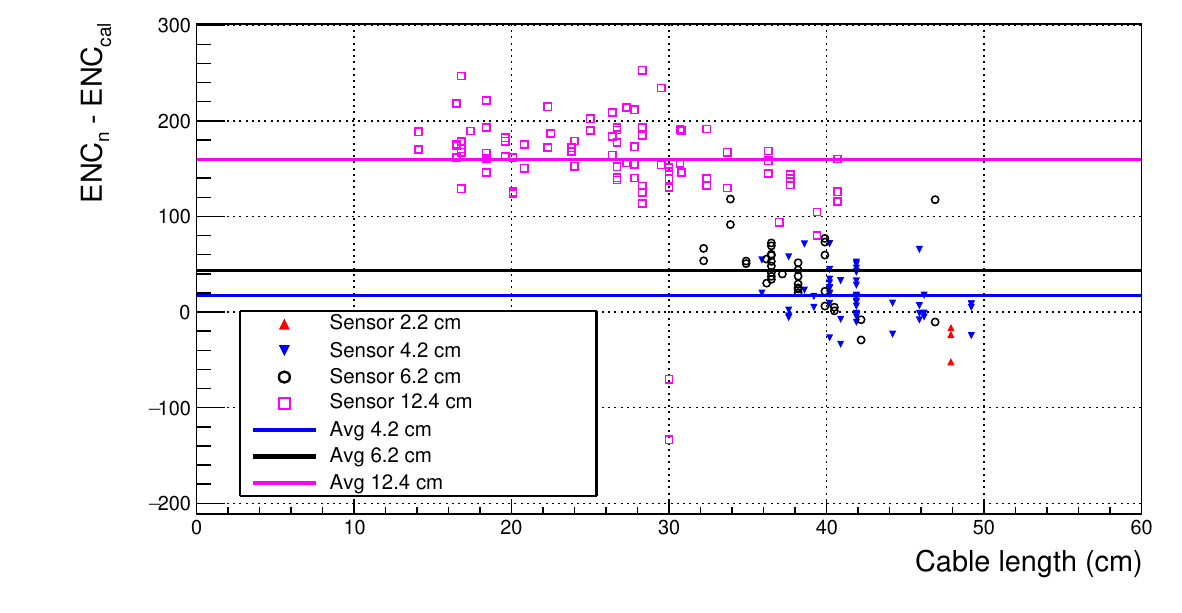}
\caption{The difference between measured and estimated ENC for n-side of modules of different form factors.}
 \label{fig:enc-tot}
\end{figure}

The measured noise can be compared to the theoretical estimate of thermal noise, based on the total capacitive load, using the expression from Ref.~\cite{Rodriguez2024}. The difference between the observed and estimated ENC for various module form factors is presented in Fig.~\ref{fig:enc-tot}. For larger sensor sizes, the ENC values appear to be underestimated, with the exact cause still under investigation.

\section{Detector prototypes and applications}
\paragraph{The mSTS setup at SIS18}

Over the past decade, the STS detector has undergone multiple prototyping iterations and beam test campaigns. These efforts culminated in the establishment of a permanent mini-CBM beam test setup at the SIS18 accelerator at GSI, with mSTS serving as the key tracking detector \cite{Dvorak2023}. More details on the mSTS performance and operation can be found in Ref.\,\cite{RamirezZaldivar2023, RamirezZaldivar2024}.

\begin{figure}[h]\centering
 \includegraphics[width=.85\linewidth]{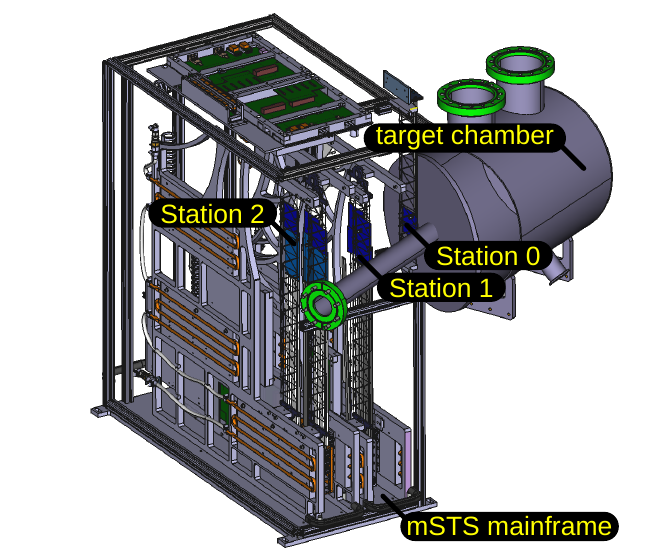}
\caption{CAD rendering of the mSTS setup.}
 \label{fig:msts}
\end{figure}

\begin{figure}[h]\centering
 \includegraphics[width=.8\linewidth]{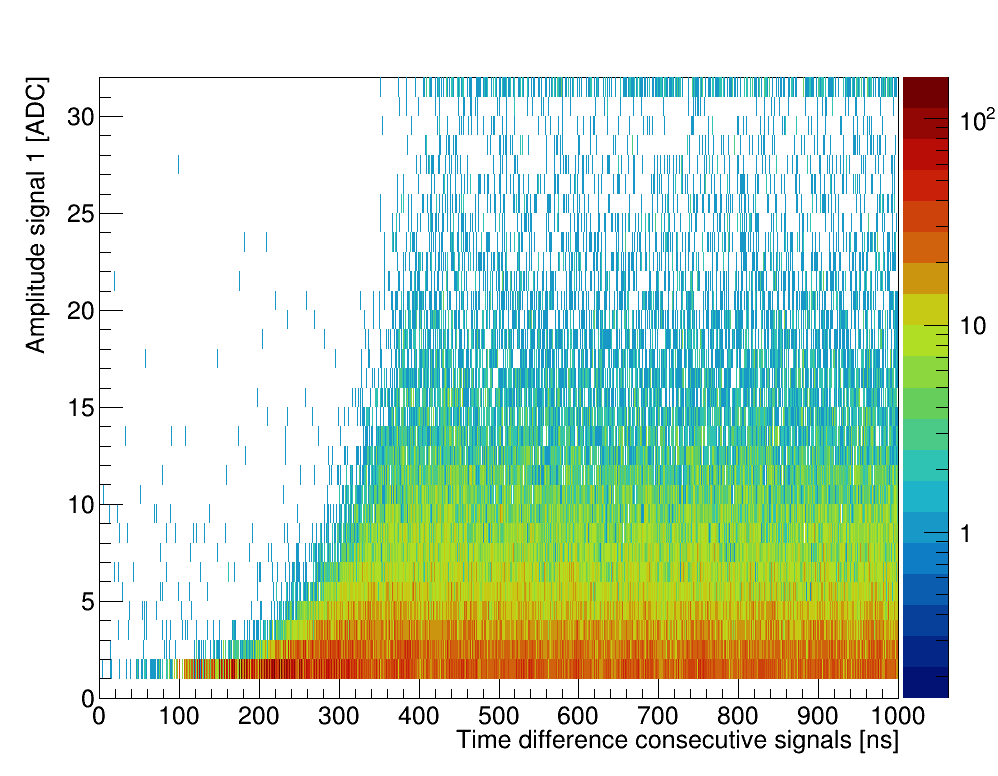}
\caption{Distribution of the time interval between sequential signals as a function of the amplitude of the initial signal.}

 \label{fig:deadtime}
\end{figure}

High-intensity heavy-ion beams from SIS18 enabled rate scan studies, allowing us to evaluate detector performance under high-occupancy conditions. The distribution of the time interval between sequential signals, relative to the amplitude of the initial signal, as shown in Fig.~\ref{fig:deadtime}, suggests a single-channel dead time of $\approx 350\,\mathrm{ns}$.

\paragraph{E16 experiment at J-PARC}
Ten STS modules with $62\times62\,\mathrm{mm^2}$ sensors are currently deployed as the first layer of the silicon tracker for the E16 experiment at J-PARC, Japan. Details on detector integration and performance during recent beam campaigns are summarized in Ref.~\cite{Aoki_2025}.

\paragraph{Beam monitoring for hadron therapy}
We recently fabricated a prototype reaction product tracking system with the potential to monitor energy deposition in hadron and heavy-ion beam cancer therapy. The current setup consists of two $62\times62\,\mathrm{mm^2}$ STS modules housed in a light-tight aluminum enclosure, with preliminary beam tests currently underway \cite{DPG2025}.

\bibliography{mybibfile}

\end{document}